\documentclass[preprint,showpacs,preprintnumbers,amsmath,amssymb]{revtex4}
\usepackage{graphicx}% Include figure files

\begin{document}

\title{Unidirectional rotary nanomotors \\
powered by an electrochemical potential gradient}
\author{ A. Yu. Smirnov$^{1,2}$, S. Savel'ev$^{1,3}$,  L. G. Mourokh$^{1,4}$, and Franco Nori$^{1,5}$ }

\affiliation{  Frontier Research System, The Institute of Physical
and
Chemical Research (RIKEN), \\
Wako-shi, Saitama, 351-0198, Japan \\
$^2$ CREST, Japan Science and Technology Agency, Kawaguchi,
Saitama, 332-0012, Japan \\
$^3$ Department of Physics, Loughborough University, Loughborough
LE11 3TU, UK \\
$^4$ Department of Engineering Science and Physics, College of
Staten Island, The City University of
New York, Staten Island, New York 10314, USA \\
$^5$ Center for Theoretical Physics, Physics Department, The
University of Michigan, Ann Arbor, MI 48109-1040, USA}

\date{\today}

\begin{abstract}
{We examine the dynamics of biological nanomotors within a simple
model of a rotor having three ion-binding sites. It is shown that in
the presence of an external dc electric field in the plane of the
rotor, the loading of the ion from the positive side of a membrane
(rotor charging) provides a torque leading to the motor rotation. We
derive equations for the proton populations of the sites and solve
these equations numerically jointly with the Langevin-type equation
for the rotor angle. Using parameters for biological systems, we
demonstrate that the sequential loading and unloading of the sites
lead to the unidirectional rotation of the motor. The previously
unexplained phenomenon of fast direction-switching in the rotation
of a bacterial flagellar motor can also be understood within our
model. }
\end{abstract}


\maketitle

Biological rotary motors provide remarkable examples of how
electrochemical energy can be efficiently converted into mechanical
motion \cite{Alberts02,BrowneNat06}. Two of the most important
representatives of this family, the F$_0$ motor of ATP (adenosine
triphosphate) synthase (F$_1$F$_0$--ATPase) and the bacterial
flagellar motor (BFM), are powered by H$^+$ or Na$^+$ ions, which
flow down the electrochemical gradient across the mitochondrial or
cell membranes, thereby generating a torque \cite{Oster98,Berg03}.
Hereafter we concentrate on proton-driven motors, but the same
considerations are valid for sodium-driven motors as well. The
gradient of the electrochemical potential is maintained by the
metabolic mechanism translocating protons from the negative side of
the membrane to its positive side \cite{Alberts02,Pumps}.

Both of these rotary motors have similar components: (i) a stator,
tightly attached to the membrane, and (ii) a ring-shaped rotor,
which can freely rotate around its axis. The rotor part of the F$_0$
motor is mechanically coupled to the F$_1$ domain of ATP synthase,
whereas the rotor of the BFM is linked to the propeller. It is
assumed \cite{Oster98} that the rotor has several (10 to 14)
proton-binding sites. The permanent generation of the torque can be
derived from the electrostatic interaction between stator charges
and charges of the rotor sites which bias the thermal diffusion of
the rotor in a specific direction \cite{Oster98,Berry00}. This
Brownian ratchet mechanism can generate a torque of about 40
pN$\cdot$nm corresponding to a realistic rate of ATP synthesis
\cite{Oster98}. However, both the power stroke and Brownian ratchet
models, exploiting ``constructive" features of Brownian motion, do
not succeed in explaining the amazing performance of the bacterial
flagellar motor, generating a torque of 2700 - 4600 pN$\cdot$nm with
an efficiency of about 90\% \cite{Berg03}. The ability of the BFM to
rapidly switch the direction of the rotation remains unexplained as
well, because transport in ratchets (see, e.g., the reviews in
\cite{DMLRatchets}) is usually fully controlled by a {\it fixed }
asymmetry of the potential energy. These facts point to the
possibility that the BFM can use the energy stored in the proton
electrochemical gradient {\it directly}, without a mediation of the
Brownian motion \cite{Berry00}.

Here we explore a simple model mimicking important features of real
biomolecular rotary motors and allowing a quantitative treatment
based on methods of condensed matter physics \cite{Wingr93}. These
approaches have been previously applied to nanoelectromechanical
systems (NEMS) with their mechanical motion affecting the electrical
properties of  electronic devices \cite{NEMS}. Similar processes
take place in nanoscale biological objects, where electrical and
mechanical degrees of freedom are also strongly coupled, making them
living counterparts to artificial NEMS. Note that only
nano-oscillators have been extensively studied by theorists,
although a single-molecule rotor
 and a nanoelectromechanical rotational actuator
\cite{GimZet} have been demonstrated experimentally. To the best of
our knowledge, no theoretical investigations of rotary NEMS have
been reported yet.

\textit{Model.} The system under consideration here consists of {\it
three} equally-spaced proton-binding sites $A,B,C$ attached to the
rotating ring (rotor) in the presence of a constant $y$-directed
electric force ${\textbf F}$ (see the inset in Fig.1). The rotor
sites can be coupled to two proton sources $S_1$ and $S_2$ as well
as to the proton drain $D$, connected to the proton reservoir with a
low electrochemical potential $\mu_D$ (the negative side of the
membrane). The source leads $S_1$ and $S_2$ can be connected (or
disconnected) at will to the proton reservoir with a higher
electrochemical potential $\mu_S$ (the positive or P-side of the
membrane). We show below that the activation of the lead $S_1$
results in a clockwise motion of the rotor, whereas connecting the
$S_2$-lead to the P-side of the membrane (and disconnecting the
$S_1$-lead) generates a counterclockwise rotation. At each instant
of time only one source lead is coupled to the P-side proton
reservoir.

The Hamiltonian of the system has the form:
\begin{eqnarray}
H = \frac{p^2}{2 M r_0^2} - U_0 \sum_{\sigma} n_{\sigma} \cos(\phi +
\phi_{\sigma}) + \sum_{\sigma} E_{\sigma} n_{\sigma} +
\nonumber\\
\sum_{k\alpha} E_{k\alpha} c_{k\alpha}^+c_{k\alpha} + \sum_{\sigma
\sigma'} U_{\sigma \sigma'} n_{\sigma} n_{\sigma'} + H_{\rm tun} +
H_{Q}, \label{H0}
\end{eqnarray}
where $\phi$ is the angle of rotation, $p = -i\hbar
(\partial/\partial \phi)$ is the operator of angular momentum of the
rotator with radius $r_0$ and effective mass $M$. We take into
account here the effects of a constant $y$-directed external
electric field, with a potential energy profile  $U(\phi) = - U_0
\cos \phi$, on the protons localized in the three sites $\sigma =
A,B,C$ with positions, characterized by the angles
$\phi_A,\phi_B,\phi_C,$ respectively. The operators $a_{\sigma}^+,
a_{\sigma}$ describe the creation and annihilation of a proton on
the site (dot) $\sigma$ with a population $n_{\sigma} =
a_{\sigma}^+a_{\sigma}$, whereas the operators
$c_{k\alpha}^+,c_{k\alpha}$ are related to the $k$-state of the
proton in the source and drain reservoirs (leads) with energy
$E_{k\alpha} \ (\alpha = S_1,S_2,D)$. The Coulomb repulsion between
protons is given by the potentials $U_{\sigma \sigma'}.$ The
tunneling coupling between dots and leads is given by the
Hamiltonian $$ H_{\rm tun} = - \sum_{k\alpha \sigma}
T_{k\alpha}c_{k\alpha}^+ a_{\sigma} w_{\alpha \sigma}(\phi) + h.c.,
$$ where the tunneling amplitudes $T_{k\alpha}$ are multiplied by the
factor $$ w_{\alpha \sigma}(\phi) = \exp\left[ -\
\frac{\sqrt{2}r_0}{\lambda} \sqrt{ 1 - \cos(\phi + \phi_{\sigma} -
\phi_{\alpha})}\right],$$ which reflects an exponential dependence
of the tunneling rate on the distance between the $\sigma-$dot and
the $\alpha-$lead with a characteristic spatial scale $\lambda$. To
take into account the influence of the classical dissipative
environment $\{ Q\}$ with the Hamiltonian $H_{\rm Bath}$ on the
rotational degrees of freedom, we include the term $H_Q = - r_0
\,\phi\, Q + H_{\rm Bath}$ in Eq.(\ref{H0}). Here $Q$ is the bath
variable, which can be represented as a sum of the fluctuating part,
$Q^{(0)}$, and the bath response on the action of the rotator, $ Q =
Q^{(0)} - \zeta r_0 \dot{\phi},$ where the coupling of the rotor to
the bath is characterized by the drag coefficient $\zeta$. The
unperturbed bath variables $Q^{(0)}$ have Gaussian statistics with
zero average, $\langle Q^{(0)}\rangle = 0$, and the correlation
function, $\langle Q^{(0)}(t)Q^{(0)}(t')\rangle = 2 \zeta T \delta
(t-t')$, where $T$ is the temperature of the environment ($k_B=1$).
The Brownian motion of the nanorotator is governed by the Langevin
equation
\begin{equation}
\ddot{\phi} + \gamma \dot{\phi} + \frac{U_0}{Mr_0^2} \sum_{\sigma}
n_{\sigma} \sin(\phi + \phi_{\sigma}) = \xi, \label{Langevin}
\end{equation}
where $\gamma$ is the damping rate of the rotator, $\gamma =
\zeta/M$, and the Gaussian fluctuation source $\xi = Q^{(0)}/(Mr_0)$
is characterized by the correlation function:  $\langle \xi(t)
\xi(t')\rangle = 2 \gamma (T/Mr_0^2) \delta(t-t').$ The influence of
the proton tunneling events, $\sim -\partial H_{\rm tun}/\partial
\phi$, on the mechanical motion is assumed to be negligibly small
here.

To describe the process of loading and unloading of proton-binding
sites $A$, $B$, and $C$, we introduce the proton vacuum state
$|1\rangle = |\rm Vac\rangle,$ jointly with seven additional
 states, $|2\rangle = a_A^+|1\rangle, \ |3\rangle =
a_B^+|1\rangle, \ |4\rangle = a_A^+a_B^+|1\rangle, \ |5\rangle =
a_C^+|1\rangle, |6\rangle = a_A^+a_C^+|1\rangle, \ |7\rangle =
a_B^+a_C^+|1\rangle,\  |8\rangle = a_A^+a_B^+a_C^+|1\rangle. $ Each
of the proton operators can be expressed in terms of operators
$\rho_{\mu \nu} = |\mu\rangle\langle \nu| \ (\mu,\nu = 1,..,8)$. In
particular, the operator $a_{\sigma}$ has the form: $ a_{\sigma} =
\sum_{\mu \nu} a_{\sigma;\mu\nu} \rho_{\mu \nu}$, with the following
non-zero matrix elements: $ a_{A;12}=a_{A;34}=a_{A;56}=a_{A;78}=1;
a_{B;13}=- a_{B;24}=a_{B;57}=-a_{B;68}=1;
a_{C;15}=-a_{C;26}=-a_{C;37}=a_{C;48}=1. $ The populations of the
dots, $n_{\sigma} = a_{\sigma}^+a_{\sigma}$, are expressed in terms
of the diagonal operators $\rho_{\mu} \equiv \rho_{\mu \mu},$ as: $
n_A = \rho_2 + \rho_4 + \rho_6 + \rho_8, \ n_B = \rho_3 + \rho_4 +
\rho_7 + \rho_8,\  n_C = \rho_5 + \rho_6 + \rho_7 + \rho_8.$ Thus,
for the protons localized on the rotor sites $A,B,C$ we obtain the
Hamiltonian $H_{ABC} = \sum_{\mu = 1}^8 \epsilon_{\mu} |\mu\rangle
\langle \mu|,$ with an energy spectrum depending on the local value
of the rotor angle $\phi$: $\epsilon_1 = 0, \ \epsilon_2 = E_A - U_0
\cos(\phi+\phi_A), \ \epsilon_3 = E_B - U_0 \cos(\phi+\phi_B),\
 \epsilon_4 =\epsilon_2 + \epsilon_3,\
 \epsilon_5 = E_C - U_0 \cos(\phi+\phi_C), \
\epsilon_6 =\epsilon_2 + \epsilon_5, \  \epsilon_7 =\epsilon_3 +
\epsilon_5, \ \epsilon_8 =\epsilon_2 + \epsilon_3 + \epsilon_5 .$ We
assume here that the characteristic time of proton tunneling to and
out of the proton-binding sites, $\gamma^{-1},$ is much shorter than
the the time scale of the rotary angle, $\langle
\dot{\phi}\rangle^{-1},$ and that the noise produced by the proton
tunneling between the sites and the source and drain contacts has
much less effect on the mechanical motion of the rotor than the
noise $\xi$ generated by the bath $\{Q\}.$ Accordingly, we can
average the stochastic Eq.(\ref{Langevin}) over fluctuations of the
proton reservoirs without averaging over the fluctuations of the
mechanical heat bath. The partially averaged proton population
$n_{\sigma}$ involved in Eq.(\ref{Langevin}) depends on the local
fluctuating value of the rotational angle $\phi$. To determine these
populations, we derive the following master equation for the proton
distribution $\rho_{\mu}$, averaged over reservoirs fluctuations,
\begin{equation}
\dot{\rho}_{\mu} + \gamma_{\mu} \rho_{\mu} = \sum_{\nu}
\gamma_{\mu\nu} \rho_{\nu}, \ \ \ \textrm{with} \label{masterEq}
\end{equation}
$$
\gamma_{\mu\nu} = \sum_{\alpha\sigma} \Gamma_{\alpha}(\phi)\{
|a_{\sigma;\mu\nu}|^2 [ 1 - f_{\alpha}(\omega_{\nu\mu})] +
|a_{\sigma;\nu\mu}|^2 f_{\alpha}(\omega_{\mu\nu})\},
$$
$\gamma_{\mu} = \sum_{\nu} \gamma_{\nu\mu},$ and
$\Gamma_{\alpha}(\phi)=\Gamma_{\alpha} |w_{\alpha\sigma}(\phi)|^2$.
Also: $\omega_{\mu\nu} = \epsilon_{\mu}-\epsilon_{\nu} \ (\hbar=1),
\Gamma_{\alpha} = 2\pi \sum_k |T_{k\alpha}|^2 \delta (\omega -
\omega_{\mu\nu})$ \cite{Wingr93}. The protons in the reservoirs are
characterized by the Fermi distributions, $f_{\alpha}(\omega) =
\left[\exp(\frac{\omega-\mu_{\alpha}}{T}) + 1\right]^{-1}$, with
temperature $T$ and electrochemical potentials $\mu_{S} = V/2,
\mu_{D} = -V/2, $ where $V$ is the proton voltage build-up. We
include the absolute value of the proton charge, $|e|$, into the
definition of the voltage $V$ and measure the voltage in units of
energy, meV. Notice that, despite of the averaging over proton
reservoirs, the master equation, Eq.(\ref{masterEq}), contains a
stochastic component, which is determined by the fluctuations of the
rotor angle $\phi(t)$, taken at the same time $t$.
Eq.(\ref{masterEq}) is valid for weak coupling between dots and
leads, and for the case when the angular coordinate $\phi(t)$ does
not change significantly on the characteristic time scale, $\tau_T =
\pi/T,$ of the Fermi distribution.

We consider here the rotation of the system with the parameters
roughly corresponding to the rotor part of the bacterial flagellar
motor \cite{Rosier98}, which has mass $M$ = 5000 kDa = 8.3 $\times
10^{-21}$ kg  and a radius $r_0 = 15 $ nm. Three torque-generating
sites,
 $A, B,$ and $C$, are attached at the points $\phi_A = 0,\ \phi_B = 2 \pi/3, $ and $ \phi_C =
- 2 \pi/3,$ respectively. The locations of the two possible source
contacts $S_1, \ S_2$ and the drain $D$ are defined as $\phi_{S_1} =
- 2\pi/3, \ \phi_{S_2} = 2\pi/3,\ \phi_{D} = 0.$  For the Coulomb
interaction between sites located a distance \ $r_0\sqrt{3}$ \
apart, in a medium with a dielectric constant $\varepsilon_r \sim
4$, we obtain: $U_{ab}\simeq U_{bc}\simeq U_{ac} \simeq $ 15 meV. We
choose an intermediate value of the drag coefficient $\zeta$ = 30 pN
s/m, which is related to the motion of a sphere with radius $r_0 =
15$ nm in an environment with a viscosity of 0.1 mPa$\cdot$s. The
rotational damping rate is quite high, $\gamma$ = 3.6 ns$^{-1}$,
which means that the free rotations of the motor come to an end
after a time interval of the order $\gamma^{-1} \sim $ 300 ps.

 \textit{Results}. In Fig.1a we present a schematic diagram of the rotor ring, jointly with the time
 dependence of the number of full rotations $\phi(t)/2\pi$,
 obtained from the numerical solution of the
Langevin equation, Eq.(\ref{Langevin}), and the master equations,
Eqs.(\ref{masterEq}), at the source-drain voltage $V = \mu_S -
\mu_D$ = 500 meV, the external potential $U_0$ = 200 meV, the
tunneling couplings to the leads $\Gamma_L = \Gamma_R = 10^9 \
$s$^{-1},$ and at temperature $T$ = 300 K. The source contact $S_1$
is activated, and, accordingly, the rotor moves in the clockwise
(positive) direction and performs more than four full rotations in
$\sim$10 $\mu$s. We start here with a slightly shifted initial
position of the rotor, when $\phi(0) = - \pi/20, \ \dot{\phi}(0) =
0,$ and the sites $A$ and $C$ are displaced from the drain $D$ and
source $S_1$ contacts, respectively. The pronounced influence of the
environment on the rotational degrees of freedom is reflected in the
noisy time dependence (see Fig.1b) of the speed of rotations
$\Omega(t)/2\pi,$ where $\Omega(t) = \dot{\phi}(t).$ Figure 1c
illustrates the synchronous dynamics of loading and unloading the
proton binding sites $A, B,$ and $C$ having populations $n_A,\ n_B,$
and $ n_C,$  when they pass through the source and the drain leads.
The site $C$ is populated first because of its close proximity to
the source lead $S_1$. The external electric field pushes the
$C$-proton, and, correspondingly, the whole rotor unit, to turn
through the angle $2\pi/3$ to the position of the minimum of the
potential $U = - U_0 \cos(\phi + \phi_{C})$. At this position of the
rotor, the site $C$ approaches the drain contact $D$ and unloads the
proton. At the same rotor position, the site $B$ starts to be
populated since this site is in the loading range of the source lead
$S_1$. This leads to a subsequent 120$^{\circ}$-turn of the rotor.
The process repeats over and over, resulting in a continuous
unidirectional rotation of the rotary ring. The torque, exerted by
the nanomotor \cite{Berry00}, can be defined as: $ {\cal T} =
-\sum_{\sigma} \left< n_{\sigma} \ \frac{d U(\phi +
\phi_{\sigma})}{d\phi} \right>,$ where $U(\phi) = - U_0 \cos \phi$
 is the potential of the $y$-directed external electric field ${\textbf
F} = F {\textbf y}, U_0 = F r_0.$ As follows from Fig.1d the torque
oscillates between 10 and 30 pN$\cdot$nm.  These numbers are
comparable with the torque produced by the F$_0$-motor of ATP
synthase, but much less than the BFM torque. The torque of the motor
can be increased by increasing the number of torque-generating
sites.

In Fig.2  we present the dependence of the average speed of
rotations, $\langle \Omega\rangle/2\pi = \langle
\dot{\phi}\rangle/2\pi$ (left axis), and the average particle
current, $I = -I_L = - (d/dt)\sum_k \langle c_{kL}^+c_{kL}\rangle $
(right axis), on the difference of proton electrochemical potentials
of the positive ($\mu_S$) and the negative ($\mu_D$) side of the
membrane (proton voltage build-up, $V$). At the fixed voltage $V$
the average speed of rotation is linearly proportional to the
amplitude of the external electric potential $U_0$, up to the
threshold value $ U_{0,{\rm max}} = V/2$.

To demonstrate that the unidirectional rotation of the motor has an
origin which is different from the mechanisms utilizing Brownian
motion of the rotary ring, we drop the fluctuation source $\xi$ in
Eq.(\ref{Langevin}) and calculate the time dependence of the angle
$\phi$ in the absence of noise (it corresponds to the mean-field
approximation for rotational degrees of freedom). To show the
reversibility of rotations, in Fig.3 we switch also the direction of
rotation from clockwise to counterclockwise by connecting the lead
$S_2$ to the proton reservoir with a higher electrochemical
potential (and disconnecting the lead $S_1$). The initial condition
for the rotor angle remains the same, $\phi(0) = - \pi/20, \
\dot{\phi}(0) = 0. $ This shift of the rotor position hampers the
initial loading of the site $B$ from the source $S_2$ (see Fig.3c),
postponing the beginning of the full-scale revolutions.
Nevertheless, without noise, the rotor makes more than three full
rotations in 10 $\mu$s, despite the unfavorable initial conditions
and despite the strong enough damping rate $\gamma$ = 3.6 ns$^{-1}$.
Note that the time dependence of the instantaneous speed of
rotations $\Omega /2\pi$ (see Fig.3b) shows regular oscillations,
compared with the noisy time dependence of the same variable
depicted in Fig.1b. It is evident from Fig.3b that the speed of
rotations oscillates synchronously with the torque generated by the
motor (see Fig.3d).

Turning back to the situation with noise, we now estimate the
efficiency of the nanomotor proposed above. For the parameters $U_0$
= 200 meV, and $V$ = 500 meV, the average torque $\langle {\cal
T}\rangle $ is about 20 pN$\cdot$nm, while the average speed of
rotations $\langle \Omega \rangle /2\pi $ is near 420 kHz (see
Fig.1d and Fig.2). This means that for the output power of the motor
\cite{Berg03} we obtain: $\langle {\cal T}\rangle \cdot \langle
\Omega \rangle \simeq 3.3 \cdot 10^5 \ $eV/s. To calculate the input
power provided by the proton reservoirs, we assume that the proton
flux (proton current) is about $ I = 1.2 \cdot 10^6$ particles per
second (see Fig.2), and the voltage drop between the reservoirs is
$V$ = 0.5 eV. Thus, the input power can be evaluated as $ I \cdot V
= 6 \cdot 10^5 $ eV/s. Therefore, for the efficiency of the motor, $
\textrm{Eff} = \langle {\cal T}\rangle \cdot\langle \Omega
\rangle/(I \cdot V)$, we obtain the estimate: $\textrm{Eff} \simeq
55 \%$. Increasing the number of protonable sites and optimizing the
system's design and parameters would improve the performance of our
simple model, approaching the performance of real rotary biomotors,
having an efficiency near 90\%.

In summary, we have proposed a simple physical model of ion-driven
rotary nanomotors which transmit the energy of an external constant
electric field into mechanical motion in the presence of an
electrochemical gradient. This model describes the main properties
of both the F$_0$-motor of ATPase in mitochondria membranes and the
bacterial flagellar motor \textit{ without} resorting to Brownian
motion. The proposed mechanism of torque generation is based on the
subsessive loading and unloading of ion-binding sites on the rotor
in the presence of a dc electric field. Our model can explain the
previously mysterious phenomenon of rotational direction switching
which occurs in bacterial flagellar motors. Solving jointly the
Langevin-type equation and master equations for the populations of
the ion-binding sites, we have determined the time evolution of the
system for various values of voltages and amplitudes of the dc
external electric field.

This work was supported in part by the National Security Agency,
Laboratory of Physical Sciences, Army Research Office, National
Science Foundation grant No. EIA-0130383, and JSPS CTC Program.
S.S. acknowledges support from the EPSRC ARF No. EP/D072581/1 and
AQDJJ network-programme.

\newpage

\begin{figure}[ht]
\includegraphics[width=15.0cm]{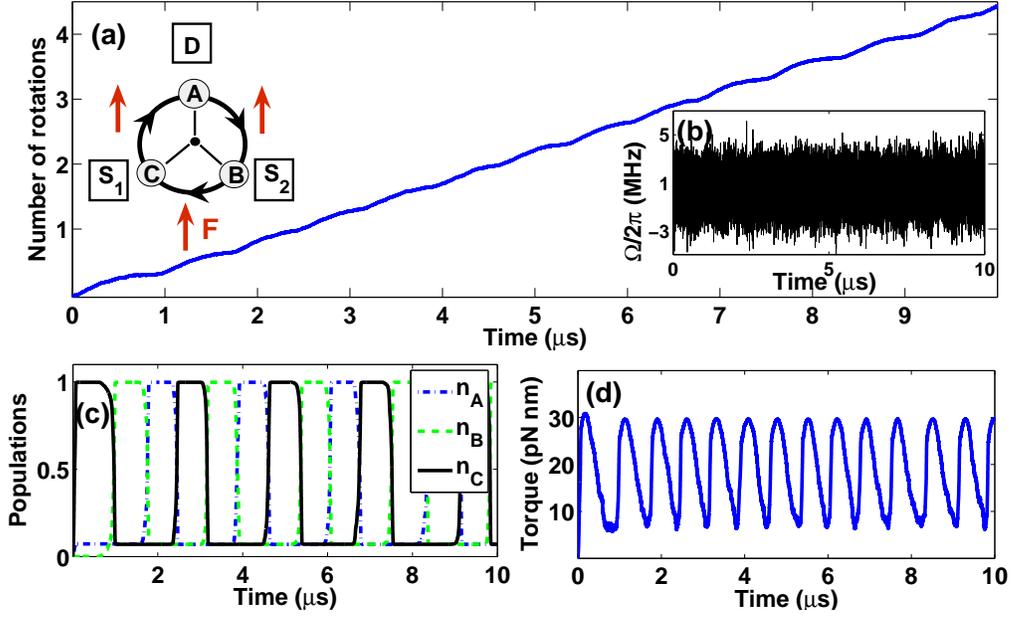}
\vspace*{3cm} \caption{ (Color online) (a) Time dependence of the
number of rotations $\phi (t) /2\pi$ at $V$ = 500 meV, $U_0$ = 200
meV, and $T$ = 300 K, with a schematic diagram of the system in the
inset; (b) Stochastic dynamics of the instantaneous speed of
rotation, $\Omega(t)/2\pi = \dot{\phi}(t)/2\pi;$ (c) Populations of
the proton-binding sites versus time; (d) Time dependence of the
rotor torque in the presence of a heat bath. Notice the periodicity
in (c) and (d).}
\end{figure}

\begin{figure}
\includegraphics[width=15.0cm]{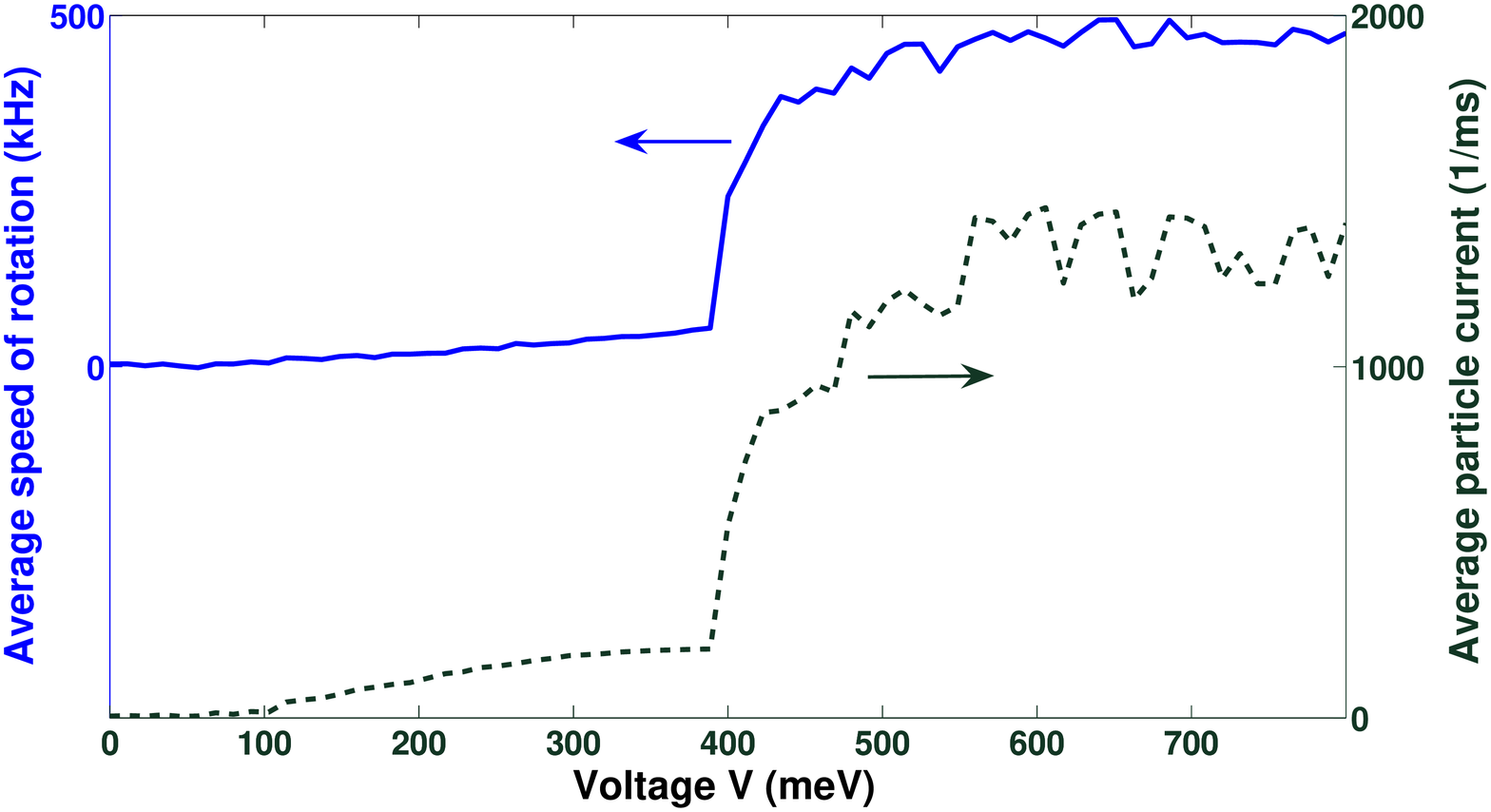}
\vspace*{1cm} \caption{(Color online) Dependence of the average
speed of rotation, $\langle \Omega\rangle /2\pi$ (solid blue curve,
left axis), and the average proton current, $I$ (dashed black curve,
right axis), on the proton voltage build-up, $V = \mu_S - \mu_D$, at
$U_0 = 200$ meV and $T$ = 300 K. }
\end{figure}

\begin{figure}
\includegraphics[width=15.0cm]{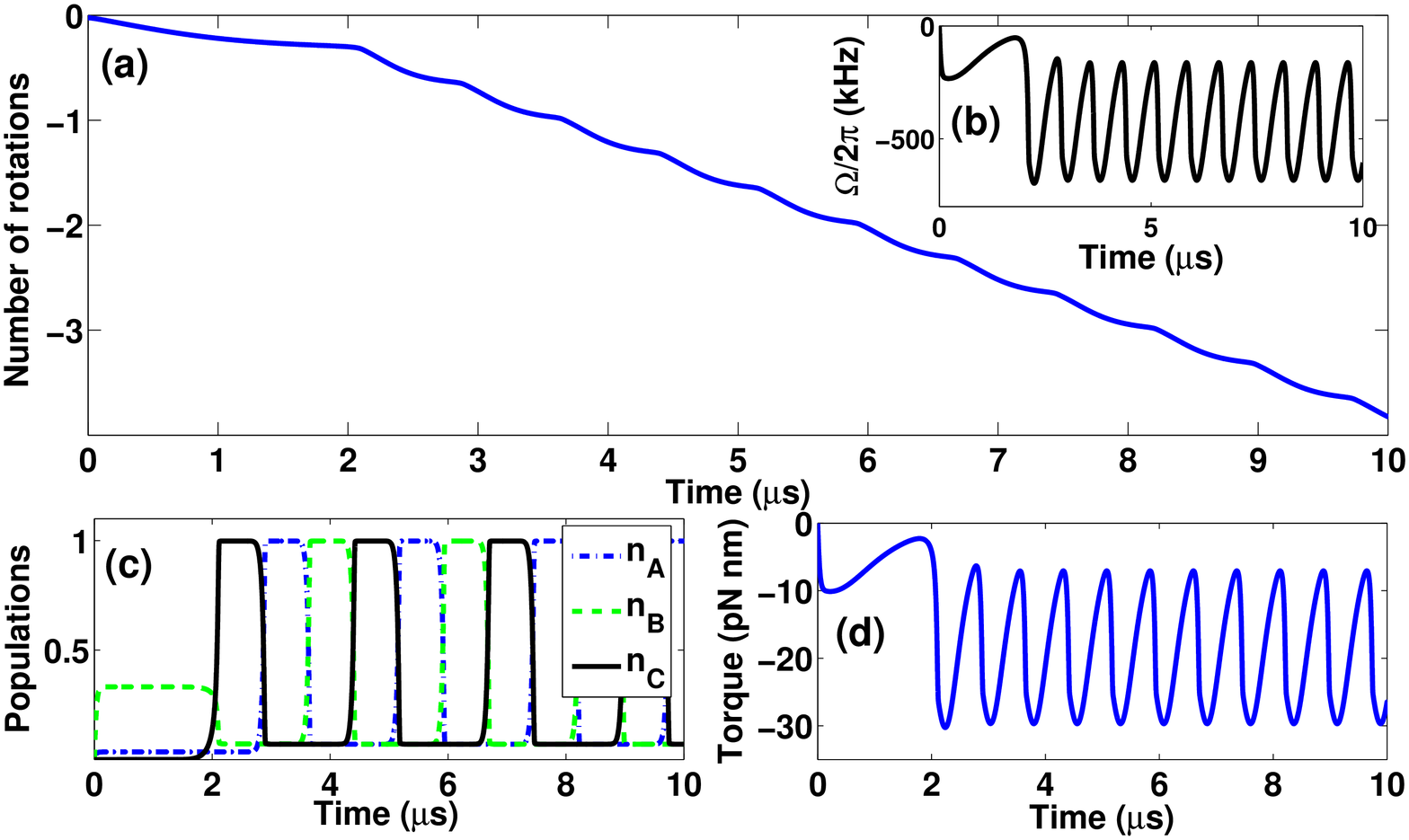}
\vspace*{1cm} \caption{(Color online) Time evolution of the system
in the \textit{absence} of external noise at $V$ = 500 meV and $U_0$
= 200 meV, but now  \textit{switching} the direction of rotations:
(a) number of full rotations, $\phi(t) /2\pi,$ as a function of
time; (b) speed of rotations, $\Omega(t)/2\pi$; (c) populations
$n_A(t), n_B(t),$ and $n_C(t)$, describing the loading and unloading
of the sites; (d) torque exerted by the rotor versus time. }
\end{figure}

\end{document}